\documentclass[twocolumn]{svjour3}                     
\smartqed  
\usepackage{graphicx}
\newcommand{\s}{~}

\graphicspath{{./figures/}}
\begin{document}

\title{ Crystal structure of Cu-Sn-In alloys around the $\eta$ phase field studied by neutron diffraction}
%

\titlerunning{Crystal structure of Cu-Sn-In alloys (...)}        

\author{ G Aurelio \and S A Sommadossi \and G J Cuello}


\institute{G Aurelio \at
               Consejo Nacional de Investigaciones Cient\'{\i }ficas y
T\'{e}cnicas, Centro At\'{o}mico Bariloche - Comisi\'{o}n Nacional
de Energ\'{\i}a At\'{o}mica, Av. Bustillo 9500, 8400 S. C. de
Bariloche, RN, Argentina \\
              Tel.: +54-294-4445180\\
              \email{gaurelio@cab.cnea.gov.ar}           
           \and
           S A Sommadossi \at
              IDEPA --Consejo Nacional de Investigaciones Cient\'{\i }ficas y
T\'{e}cnicas --Facultad de Ingenier\'ia, Universidad Nacioanl del Comahue, Buenos Aires 1400, 8300 Neuqu\'en, Argentina
\and
          G J Cuello   \at
              Institut Laue Langevin, F-38042 Grenoble, France
}

\date{Received: date / Accepted: date}

\maketitle

\begin{abstract}

The study of the Cu-Sn-In ternary system has become of great importance in recent years, due to new environmental regulations forcing to eliminate the use of Pb in bonding technologies for electronic devices. 
A key relevant issue concerns the intermetallic phases which grow in the bonding zone and are determining in their quality and performance. In this work, we focus in the $\eta$-phase (Cu$_2$In or Cu$_6$Sn$_5$) that exists in both end binaries and as a ternary phase. We present a neutron diffraction study of the constitution and crystallography of a series of alloys around the 60 at.\% Cu composition, and with In contents ranging from 0 to 25 at.\%, quenched from 300$^{\circ}$C. The alloys were characterized by scanning electron microscopy, probe microanalysis and high-resolution neutron diffraction. The Rietveld refinement of neutron diffraction data allowed to improve the currently available model for site occupancies in the hexagonal $\eta$-phase in the binary Cu-Sn as well as in ternary alloys. For the first time, structural data is reported in the ternary Cu-Sn-In $\eta$-phase as a function of composition, information that is of fundamental technological importance as well as valuable input data for ongoing modelisations of the ternary phase diagram.

\end{abstract}

\section{Introduction \label{Introduction}}

Since the recent emergence of environmental regulations forcing to eliminate the use of Pb in soldering and bonding technologies from the electronic industry, the study of Pb-free alternative alloys has received major attention \cite{08Din}. Not only new materials but also new soldering methods are being explored to improve the performance of joints under high--temperature working conditions, such as the so-called Transient Liquid
Phase Bonding (TLPB) or diffusion soldering \cite{04Gal}. This method relies on a diffusion-reaction process
between an interlayer solder and the parent material or substrate, which in the electronic industry is mostly copper. The advantage of TLPB relies on producing joints with microstructures and mechanical properties very similar to those of the parent material, and capable of withstanding higher temperatures. 

Sn-based alloys are among the favorite Pb-alloy substitutes in bonding technology, having low melting points, low cost, good wettability, etc. In particular, the Sn-In binary system with an eutectic point at only 120$^{\circ}$C (In-48 at.\% Sn) constitutes an excellent candidate to lower the processing temperature of the joints. However, interfacial reactions at the solder/Cu interface and in the solder matrix lead
to the formation of intermetallic phases (IPs) which are often brittle and hard causing failure to the
solder joints. Therefore, if Sn-In alloys are to become solder candidates, the properties of the ternary 
Cu-Sn-In system and its IPs as constituents of microelectronic solder joints turn into a fundamental matter of research. Paradojically, the equilibrium phase diagram of the Cu-Sn-In system is still not completely determined and it remains in evaluation \cite{07Vel}, with a few experimental studies \cite{72Koe,08Lin,09Lin} and one complete thermodynamic calculation using CALPHAD \cite{01Liu}. However, even the binaries Cu-Sn and Cu-In present discrepancies regarding the crystallography of its numerous IPs. 

Of particular interest for industry and TLPB technology is the IP called $\eta$-phase, which is present both in Cu-In (Cu$_2$In) and in Cu-Sn (Cu$_6$Sn$_5$) alloys, as it is usually present in the interface Cu-solder. This phase presents polymorphic transitions with temperature: to avoid confussion between the two binaries, in the present work the high--temperature form will be called $\eta$(HT) and the low--temperature modification will be called $\eta$(LT). For Cu$_6$Sn$_5$ the $\eta$(HT)$\rightarrow \eta$(LT) occurs at 186$^{\circ}$C whereas for Cu$_2$In it would occur between 306$^{\circ}$C and 383$^{\circ}$C for some authors \cite{97Eld,03Bah}, while for others the sequence of phases is more complex \cite{72Jai,89Sub}. In Cu-Sn alloys, the LT$\rightarrow$HT transition is accompanied by a specific volume change of about 2.15\%  which is highly nocive for the performance of soldered joints as it produces cracks \cite{10Nog,10Sch}.

Despite the unclear scenario for the crystallography and stability of Cu$_2$In alloys, it has been proposed that Cu$_2$In and Cu$_6$Sn$_5$ form a continuous solid solution in the ternary phase diagram between 186$^{\circ}$C and 383$^{\circ}$C, but no further details are given about the ternary $\eta$ phase. It is important to note that the proposal of a continuous solid solution comes from microscopy observations of microstructure and by conventional XRPD which cannot yield any information on the atomic distribution of In and Sn in the proposed hexagonal crystallographic cell. We present in this paper the first neutron powder diffraction (NPD) study of ternary Cu-Sn-In alloys focusing on the crystal structure and phase stability of the $\eta$-phase. Previous works and our experience indicate that unless very carefully prepared, powders of these alloys are difficult to study using X-ray diffraction methods. This is mainly due to a large linear absorption coefficient for Cu-K$_\alpha$ radiation, and severe preferred orientation \cite{96Pep}, in particular for alloys subjected to long aging treatments. By using NPD we have greatly overcome these issues. 
The objective of the present work is to present a systematic study of the crystallography of the $\eta$ phase in ternary Cu-In-Sn alloys, tracking the atomic arrangement as the composition is varied from the binary Cu-Sn $\eta$-phase towards the binary Cu-In $\eta$-phase in the 300$^{\circ}$C isothermal section. 

\section{Experimental details \label{Exp}}

\subsection{Alloys}

Seven Cu-In-Sn alloys with nominal compositions lying in the proximity of the $\eta$ phase field are discussed in this paper. The alloys were prepared by melting nominal amounts of the high purity elements (4N) in an electric furnace under a reducing Ar atmosphere, and cooled in air. The resulting ingots were encapsulated in quartz ampoules under Ar atmosphere, annealed at $300^{\circ}$C for 3 weeks to promote homogenization, and then rapidly quenched to $0^{\circ}$C. 

Optical microscopy inspection of Sn-rich alloys S1 to S5 reveals a microstructure composed of well developped grains with an avarage size of 300\s$\mu$m (matrix), some Sn seggregation at grain boundaries, and small amounts of a second phase close to the edges with a different morphology, which tend to disappear with increasing In content. We will further discuss this issue in Section\s\ref{Results}. 

The composition of the alloys was determined by electron probe microanalysis (EPMA) using the WDS technique under an acceleration voltage of 15kV. Measurements were performed on several spots corresponding to the matrix, grain boundaries and seggregated phase. Results are summarised in Table\s\ref{t:1}. Based on the phase diagrams available for these alloys\s\cite{90Mas}, the matrix would correspond to the $\eta$-phase and the seggregated striped phase to the $\varepsilon$-phase Cu$_3$Sn, in agreement with microanalysis results.

\begin{table*}
 \caption{Composition of the Cu-In-Sn alloys studied in the present work. Samples S1 to S6 were measured by WDS, selecting grains from the middle and from each edge of the ingots. The global composition of sample S7 (marked with $^*$) was measured by EDS. }
 \begin{center}
\begin{small}
\begin{tabular}{lcccccccccccc}
Sample & \multicolumn{3}{c}{Nominal} &  \multicolumn{3}{c}{Matrix} & \multicolumn{3}{c}{Seggregated} & \multicolumn{3}{c}{Grain Boundaries} \\ 
  & \multicolumn{3}{c}{at.\%} &  \multicolumn{3}{c}{at.\%} & \multicolumn{3}{c}{at.\%} & \multicolumn{3}{c}{at.\%} \\\cline{2-13}
 & Cu & In & Sn & Cu & In & Sn & Cu & In & Sn & Cu &In &Sn\\ \hline
S1 & 55 & 0 & 45 & 53.9 &  & 46.0  & 73.8  & 0.0 & 26.1  & 3.0  & 0.2 & 96.8  \\
S2 & 55 & 1 & 44 & 54.3 & 1.1 & 44.5 & 74.2 & 0.2 & 25.6 & 2.1 & 2.1 & 95.8 \\
S3 & 55 & 3 & 42 & 53.9 & 3.1 & 43.0 & 73.7 & 0.6 & 25.6 & 2.02 & 6.07 & 91.9 \\
S4 & 55 & 5 & 40 & 55.6 & 5.0 & 39.4 & 74.0 & 1.1 & 24.9 & 0.9 &  57.2 & 87.7 \\
S5 & 58 & 12 & 30 & 58.1 & 12.2& 29.7 & 74.2 & 3.2 & 22.6 & -- & -- &  --\\
S6 & 60 & 20 & 20 & 61.2 & 20.4 & 18.4 &-- & -- & -- & -- & -- & -- \\
S7 & 60 & 24 & 16 & 60.8$^*$ & 24.5$^*$ & 14.7$^*$ &-- & -- & -- & -- & -- & -- \\
\end{tabular}
\end{small}
 \end{center}
\label{t:1}
\end{table*}

\begin{figure}[ptb]
\centering
\includegraphics[width=0.75\linewidth]{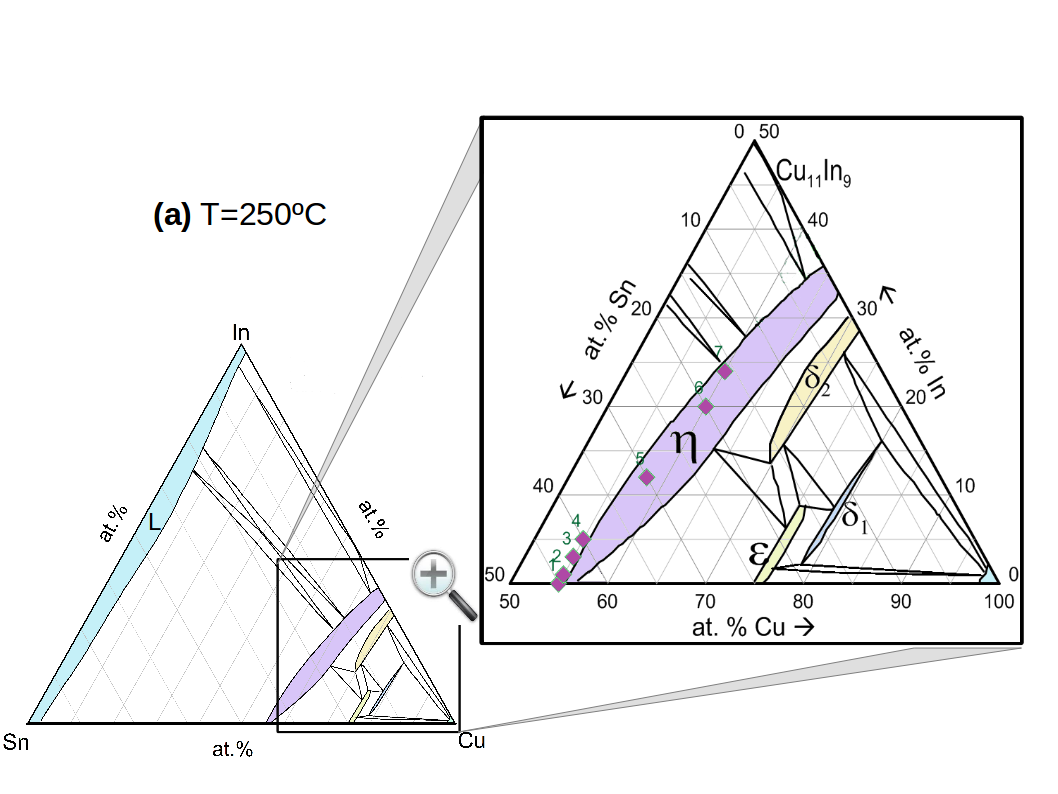}
\includegraphics[width=0.75\linewidth]{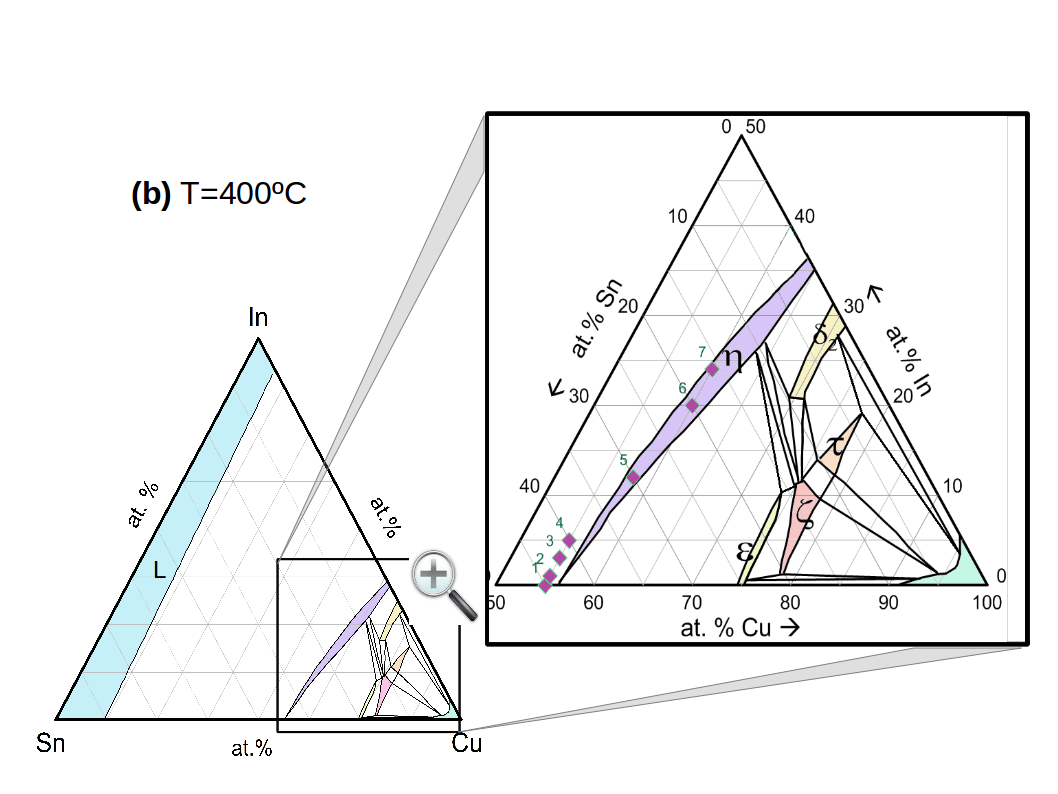}
\caption{Isothermal sections of the Sn-In-Cu ternary system at (a) $250^{\circ}$C after Lin \textit{et al}\s\cite{09Lin} and (b) $400^{\circ}$C after K\"oester \textit{et al}\s\cite{72Koe}. Symbols denote the alloys studied in the present work, while shaded areas correspond to the following phase fields: $\eta$: Cu$_6$Sn$_5$, Cu$_2$In; $\varepsilon$: Cu$_3$Sn; $\delta_1$: Cu$_{41}$Sn$_{11}$; $\delta_2$: Cu$_7$In$_3$; $\zeta$: Cu$_{10}$Sn$_3$; $\tau$: Cu$_{11}$In$_2$Sn \cite{72Koe} or Cu$_{16}$In$_3$Sn \cite{01Liu}.} \label{f:PDs}
\end{figure}

In Fig.\s\ref{f:PDs} we present the two currently available isothermal sections of the ternary phase diagram most close to the annealing temperature used in the present work: at $250^{\circ}$C \cite{09Lin} and at $400^{\circ}$C \cite{72Koe}. Superimposed to the diagrams we have indicated the nominal composition of our alloys labelled S1 to S7.

\subsection{Neutron diffraction}

Neutron powder diffraction (NPD) experiments were performed at the Institut Laue-Langevin (ILL) in Grenoble, France. High--resolution NPD data at room--temperature were collected at diffractometer
Super--D2B. A wavelength of $\sim 1.59$~{\AA} was used, with an angular span of $150{^{\circ }}$ and a step of $0.05{^{\circ}}$. The collimated beam on the sample was of 32mm x 9mm using 500mm horizontal slits. A Si standard was used to calibrate the neutron’s wavelength, yielding the value $\lambda=1.5937\pm 0.0002$ \AA. This was in very good agreement with the calibration performed by the beamline staff using a NaCaAlF standard ($\lambda=1.5935$ \AA).  

Samples for NPD experiments were obtained by manually grinding the alloys' ingots for 10 minutes in an agate mortar, resulting in a fine powder. Measurements were performed in vanadium cylinders of 6mm diameter and 8cm heigth filled with sample. Patterns were collected in 150 minutes and the diffractograms thus obtained were processed with the full-pattern analysis Rietveld method, using the program \begin{scriptsize} FULLPROF\end{scriptsize}~\cite{fullprofB}. 

\section{Results \label{Results}}

\subsection{Constitution of the alloys}

\begin{figure}[ptb]
\centering
\includegraphics[width=0.45\linewidth]{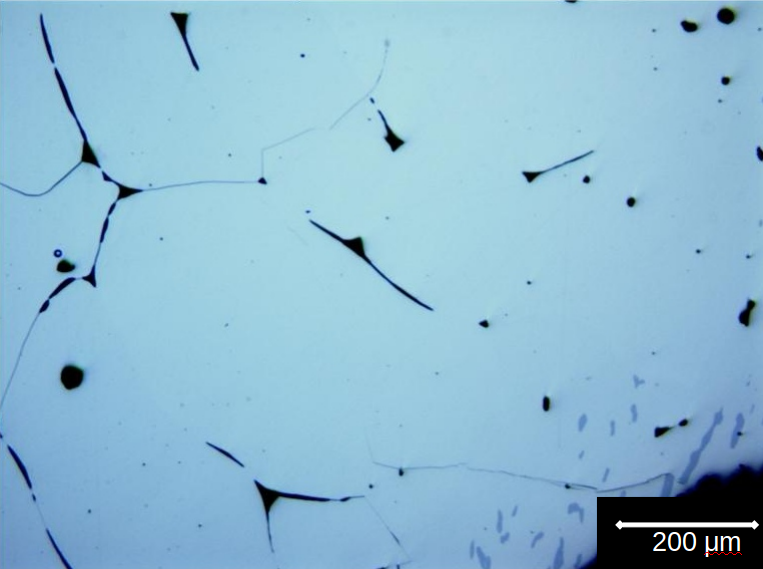} \includegraphics[width=0.45\linewidth]{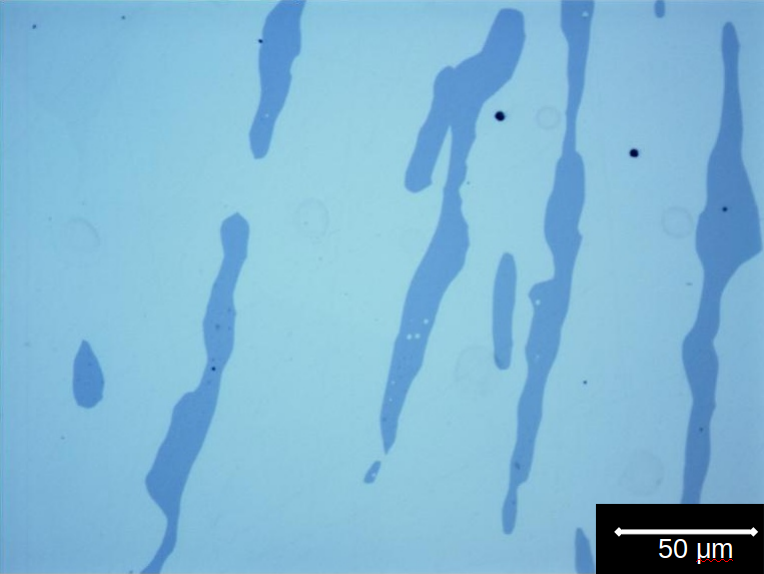}
\caption{Optical micrographs for the S1 sample at different magnifications.} \label{f:metall}
\end{figure}

Optical micrographs of sample S1 at different magnifications are presented in Fig.\s\ref{f:metall}. These images show the presence of a main homogeneous matrix with the Cu$_6$Sn$_5$ ($\eta$) composition, some segreggated Cu$_3$Sn ($\varepsilon$) with striped morphology and pure Sn at the grain boundaries. The high--resolution NPD data for this sample confirms the coexistance of the $\eta$ and $\varepsilon$ phases, but no traces were found of the crystalline $\beta$-Sn phase, suggesting that the quenching process resulted in retention of non-crystalline Sn from the melt. This is also supported by the background profile in the diffractograms. Diffraction from the $\eta$ and $\varepsilon$ phases show, on the other hand, a high degree of crystallinity. 

\begin{figure}[htb]
\centering \vspace{3mm}
\includegraphics[width=0.5\linewidth]{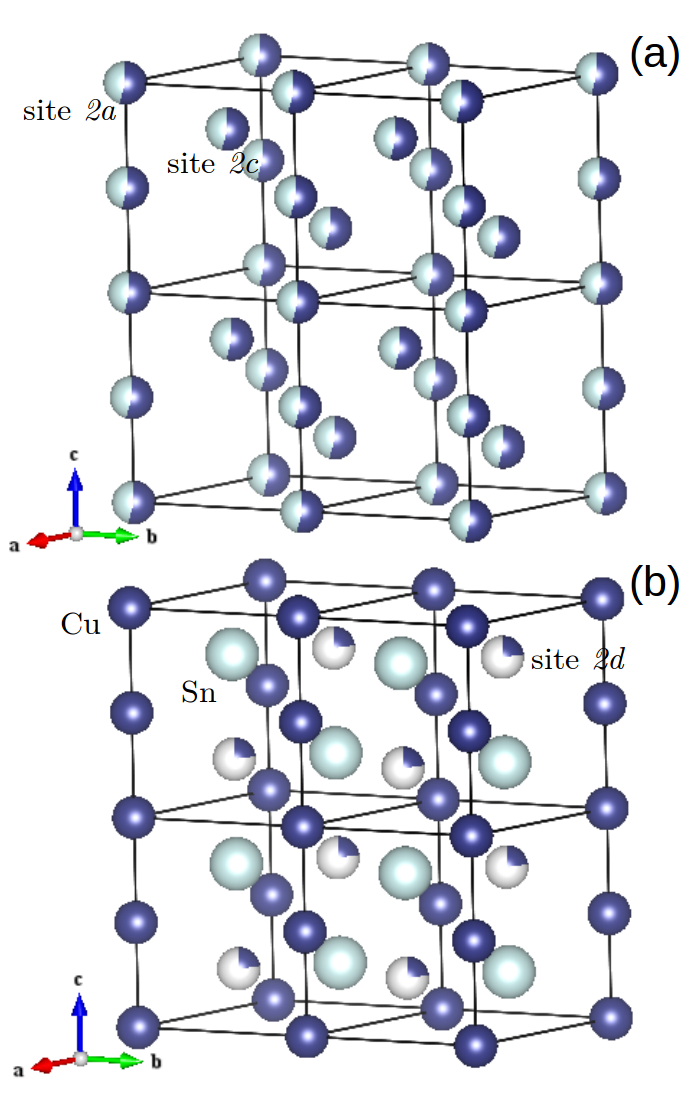}
\caption{(a) Crystallographic model currently available for the high--temperature Cu$_6$Sn$_5$ phase (ICSD 56282) in which Cu (dark atoms) and Sn (light atoms) occupy randomly the Wykoff sites \textit{2a} and \textit{2c}, represented by partial shadings. Solid lines indicate the unit cells (four in the picture) of the $P6_3/mmc$ space group. (b) Crystallographic model resulting from refinement of NPD data. Cu atoms occupy completely site \textit{2a} and In atoms site \textit{2c}. The remaining Cu partially occupies site \textit{2d}. This graph was produced using the software VESTA\s\cite{VESTA}.} \label{f:vesta}
\end{figure}

\begin{figure}[tb]
\centering \vspace{3mm}
\includegraphics[width=0.5\linewidth]{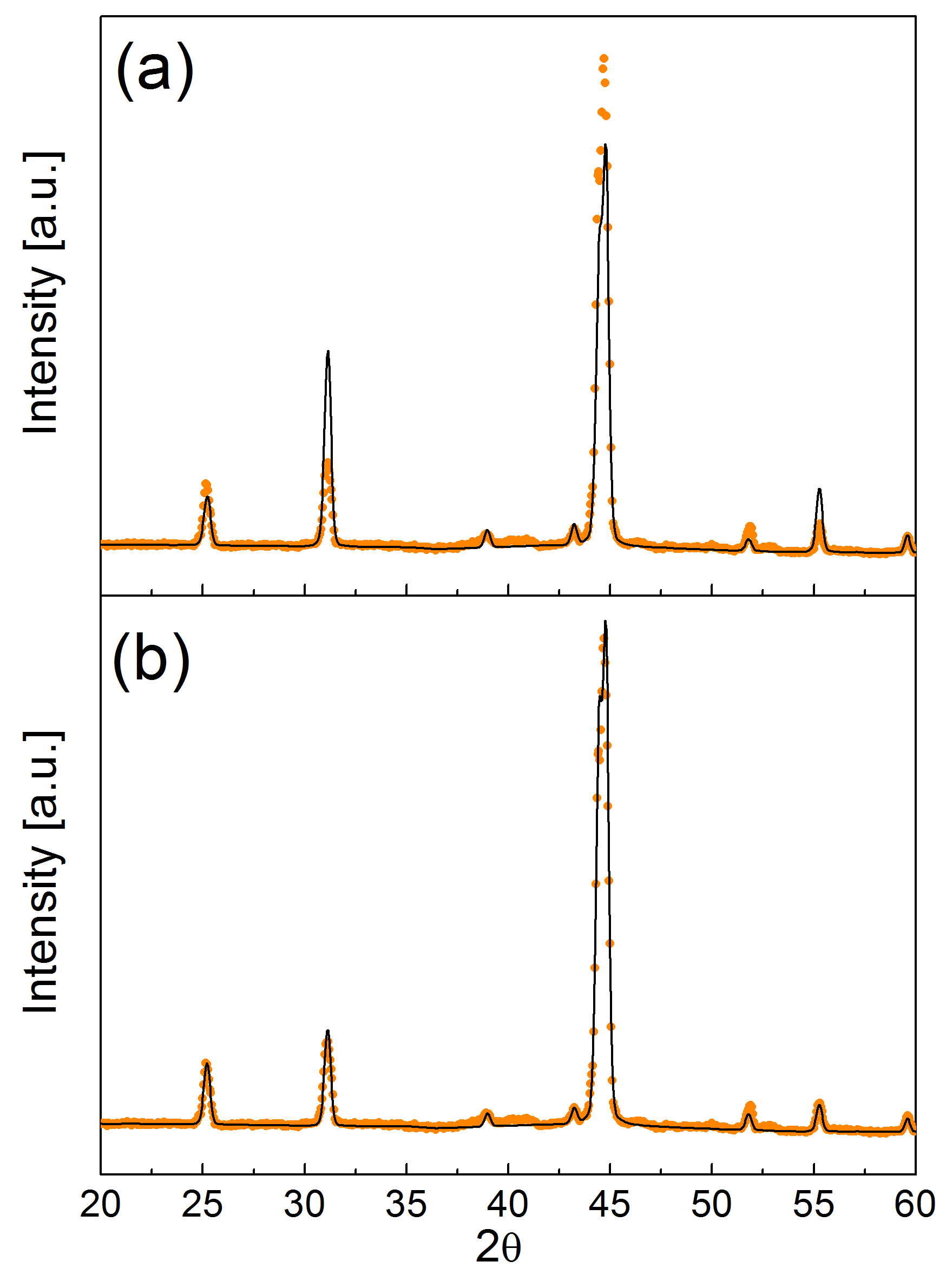}
\caption{(a) Best fit of the $\eta$-phase based on the model proposed in the databases, where Cu and Sn atoms occupy randomly the Wyckoff positions \textit{2a} and \textit{2c} of space group $P6_3/mmc$. (b) Improvement of the fit when the site occupancies correspond to the model in Fig.\s\ref{f:vesta}(b).} \label{f:compare-S1}
\end{figure}

The refinement strategy started with the available model for the high--temperature Cu$_6$Sn$_5$-phase (ICSD 56282, after ref.\cite{73Gan}) in which Cu and Sn atoms occupy randomly the Wykoff positions \textit{2a} and \textit{2c} of the $P6_3/mmc$ space group , \textit{i.e.}, the occupation of each site is 55\% Cu and 45\% Sn, as schematized in Fig.\s\ref{f:vesta}(a). Although the cell metrics was correct, this model could not account for the relative intensities in our diffractogram and led to poor fits, as shown in Fig.\s\ref{f:compare-S1}(a). Refinements improved greatly (Fig.\s\ref{f:compare-S1}(b)) when site occupancies were set to \textit{2a} with 100\% Cu and \textit{2c} to 100\%Sn. The excess Cu was allowed to occupy the \textit{2d} site as in the high--temperature Cu$_2$In $\eta$-phase (ICSD 102982 after ref.\s\cite{42Lav}, ICSD 627998 after ref.\s\cite{64Kal}). All attempts to refine these occupations with both atom types quickly converged to the same result, schematized in Fig.\s\ref{f:vesta}(b). In fact, this distribution of atoms is the one that actually corresponds to NiAs, and was also used to refine XRD data of Cu$_6$Sn$_5$ by Nogita \textit{et al}\s\cite{09Nog} as well as Peplinski \textit{et al}\s\cite{96Pep}.

\begin{figure}[htb]
\centering \vspace{3mm}
\includegraphics[width=\linewidth]{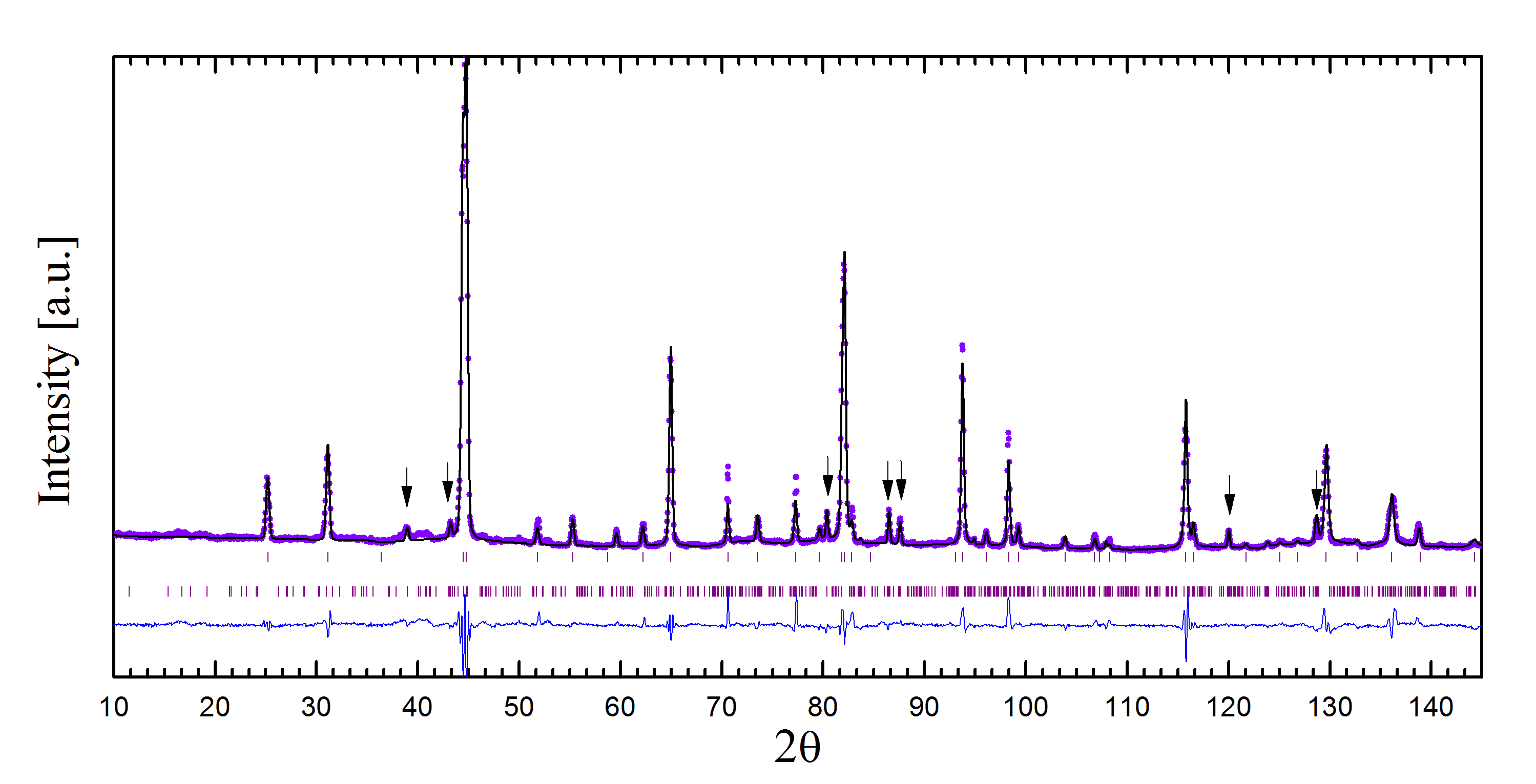}
\caption{Rietveld refinement for sample S1 with nominal 55 at.\% Cu- 45 at.\% Sn queched from $300{^{\circ}}$C, from data collected at room temperature in D2B. Vertical bars at
the bottom indicate Bragg reflections from the HT-$\eta$-phase and the orthorhombic $\varepsilon$-phase. The most intense reflections from the latter have been indicated with arrows. The line at the bottom corresponds to the difference between the experimental and calculated patterns.} \label{f:riet-S1}
\end{figure}

On the other hand, several models are available in the literature for the $\varepsilon$-phase Cu$_3$Sn, such as ICSD cards 162569, 103102, 103103, 629268. Among them, our ND data could only be successfully refined using the ortorhombic space group $Cmcm$ with a long-period superlattice (ICSD 103102 and PDF 65-5721 after \cite{83Wat}). This has been previously reported to be the case when samples are annealed for a long time in the high temperature $\varepsilon$-phase region and subsequently quenched \cite{70Bro}. As this phase is only found in samples with very little In, the refinements were performed considering stoichiometric Cu$_3$Sn to simplify the problem, although according to the available phase diagrams, this phase should present a certain In solubility too. In Fig.\s\ref{f:riet-S1} we show the complete Rietveld refinement (solid line) of the high resolution data (symbols)
collected at room temperature for sample S1. The Bragg reflections indicated at the bottom by vertical bars correspond to each of the above mentioned phases. The presence of the $\varepsilon$-phase has been highlighted using arrows. Structural data obtained from the refinement are summarised in Table\s\ref{t:2}. It is worth noting that almost all the diffractograms present a rather high background with a non-trivial dependence with 2$\theta$. The reason is most probably the presence of amorphous Sn retained by quenching from the melt. There is also a weak evidence of some LT-$\eta$-phase in very small amounts. A deeper discussion of these issues is presented in a separate report dealing with \textit{in situ} high--temperature measurements\s\cite{tobe}. For this reason, the Rietveld refinements present rather high values of $\chi^2$ and the data backgrounds are not perfectly accounted for. However, this does not affect the most relevant structural results obtained from the fit.

\begin{table*}
 \caption{Structural parameters refined from D2B data at room temperature. The $\eta-$phase was refined in the $P6_3/mmc$ space group with occupied Wyckoff positions \textit{2a}$=(0 0 0)$, \textit{2c}$=(\frac{1}{3} \frac{2}{3} \frac{1}{4})$ and \textit{2d}$=(\frac{1}{3} \frac{2}{3} \frac{3}{4})$. Only the values of refined occupations are reported in the table. Next to the refined values, the nominal occupancies for each site using model A (see text) are indicated in brackets. Numbers in itallics represent fixed parameters. The $\varepsilon-$\s phase was refined in the orthorhombic space group $Cmcm$ after Ref.\s\cite{83Wat}. Lattice parameters, phase fractions ($f$) and refined occupations are reported, as well as indicators of the quality of the fits.}
 \begin{center}
\begin{small}
\begin{tabular}{lccccccc}
\hline \hline
 & S1 & S2 & S3 & S4 & S5 & S6 & S7 \\
\textbf{$\eta$-phase} &  & & & & & & \\
 $a$ & 4.2141(1) & 4.2129(1) & 4.2157(1) & 4.2187(1) & 4.2366(1) & 4.2465(1) & 4.2500(1) \\
 $c$ & 5.1061(2) & 5.1045(2) & 5.1064(1) & 5.1091(1)& 5.1328(2) & 5.1607(2) & 5.1785(2) \\
$f_{\eta}$ & 0.95(1) & 0.98(1)& 0.99(1) & 1.00 & 0.98(2) & 1.00 & 1.00\\
$Occ_{Cu-2a}$ & 1.00 & \textit{1.00} & \textit{1.00} & \textit{1.00} & 1.00 [1.00] & 0.95 [1.00] & 1.00 [1.00] \\
$Occ_{Cu-2d}$ & 0.22 [0.22]  & 0.20 [0.22] & 0.21 [0.22]& 0.23 [0.22] & 0.35 [0.38]& 0.43 [0.5] & 0.46 [0.5] \\
$Occ_{Sn-2c}$ & 1.00 & 0.98 [0.98]& 0.955 [0.94] & 0.936 [0.89] & 0.77 [0.71] & 0.50 [0.5]& 0.40 [0.4]\\
$Occ_{In-2c}$ & 0.00 & \textit{0.02} & \textit{0.06} & \textit{0.11}& \textit{0.28} &  0.47 [0.5]& 0.61 [0.6] \\
$R_B$ & 9.44 & 8.43 & 11.6 & 9.9 & 6.8 & 14.2 & 8.7\\ \hline
\textbf{$\varepsilon$-phase} & & & & & & & \\
 $a$ & 5.5216(5) & 5.5230(5) & \textit{5.5230} & - & 5.5217(8) & - & - \\
$b$ & 47.772(5) & 47.776(5) & \textit{47.776} & - & 47.724(5) & - & - \\
 $c$ & 4.3280(5) & 4.3281(5)& \textit{4.3281} & - & 4.3302(5) & - & - \\
$f_{\varepsilon}$ & 0.05(1) & 0.02(1) & 0.01(1) & 0 & 0.02(2) & 0 & 0 \\
$R_B$ & 25.5 & 38.5 & - & - & 41 & - & - \\ \hline
$\chi^2$ & 18.4 & 30.0& 16.3 & 15.1 & 9.0 & 9.9 & 8.3\\ \hline\hline

\end{tabular}
\end{small}
 \end{center}
\label{t:2}
\end{table*}

\begin{figure}[ptb]
\centering
\includegraphics[width=0.4\linewidth]{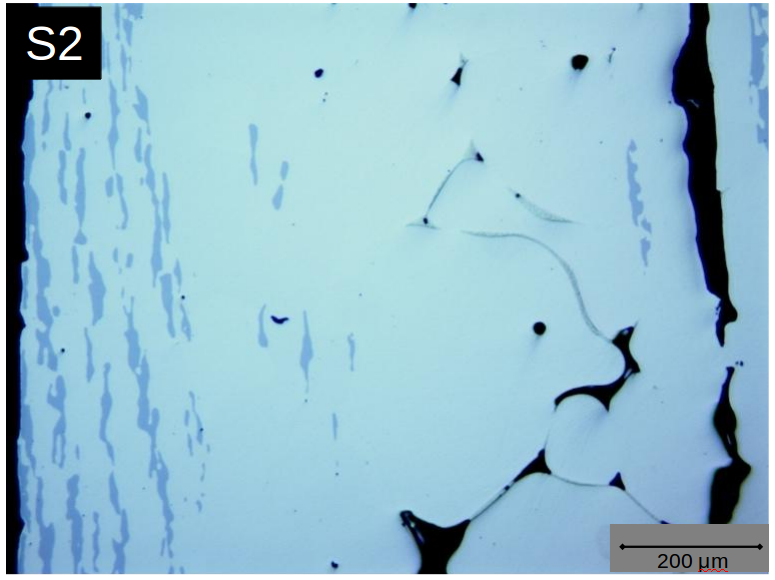}\includegraphics[width=0.4\linewidth]{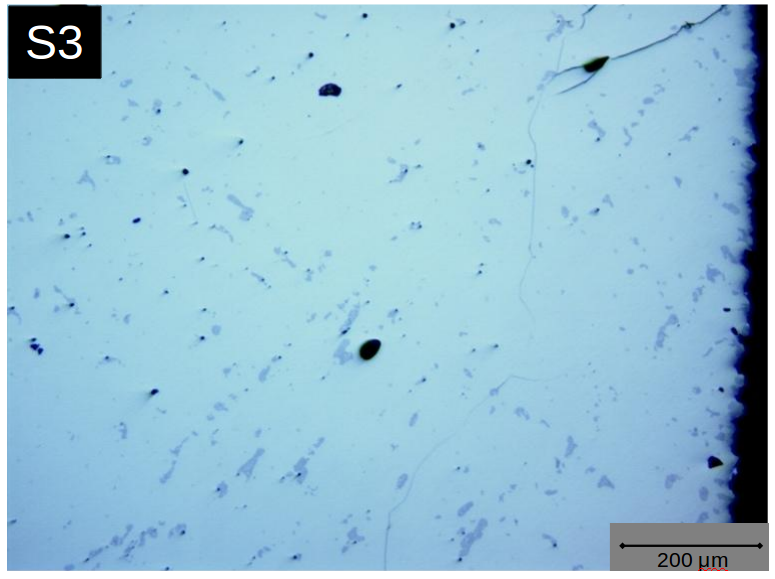}
\includegraphics[width=0.4\linewidth]{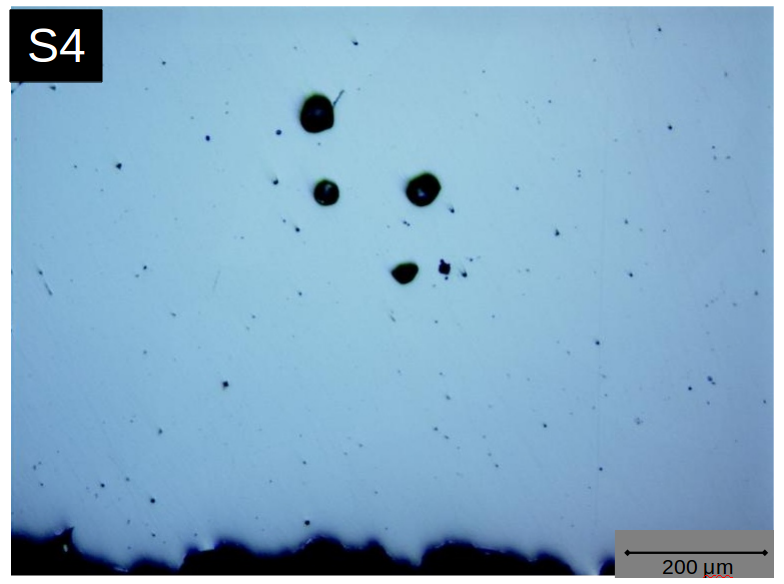}\includegraphics[width=0.4\linewidth]{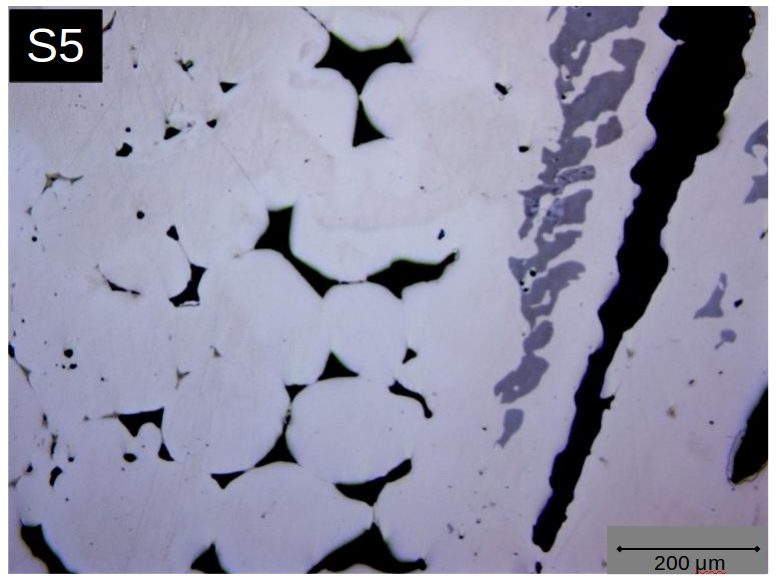}
\caption{Optical micrographs for samples S2 to S5. The $\eta$-phase grains are well developed and some $\varepsilon$-phase is also observed in most samples, particularly at the ingot's edges.} \label{f:metall-S2toS5}
\end{figure}

The same trend as in S1 is observed in the metallographies of samples with increasing In content, at least up to S5 (nominal 12 at.\% In - 30 at.\% Sn). They all show a majority homogeneous phase ($\eta$), surrounded by some grain bounday Sn-rich seggregation and a second minor seggregated phase ($\varepsilon$) mostly at the edges, as illustated in Fig.\s\ref{f:metall-S2toS5}.

\begin{figure}[tb]
\centering \vspace{3mm}
\includegraphics[width=0.75\linewidth]{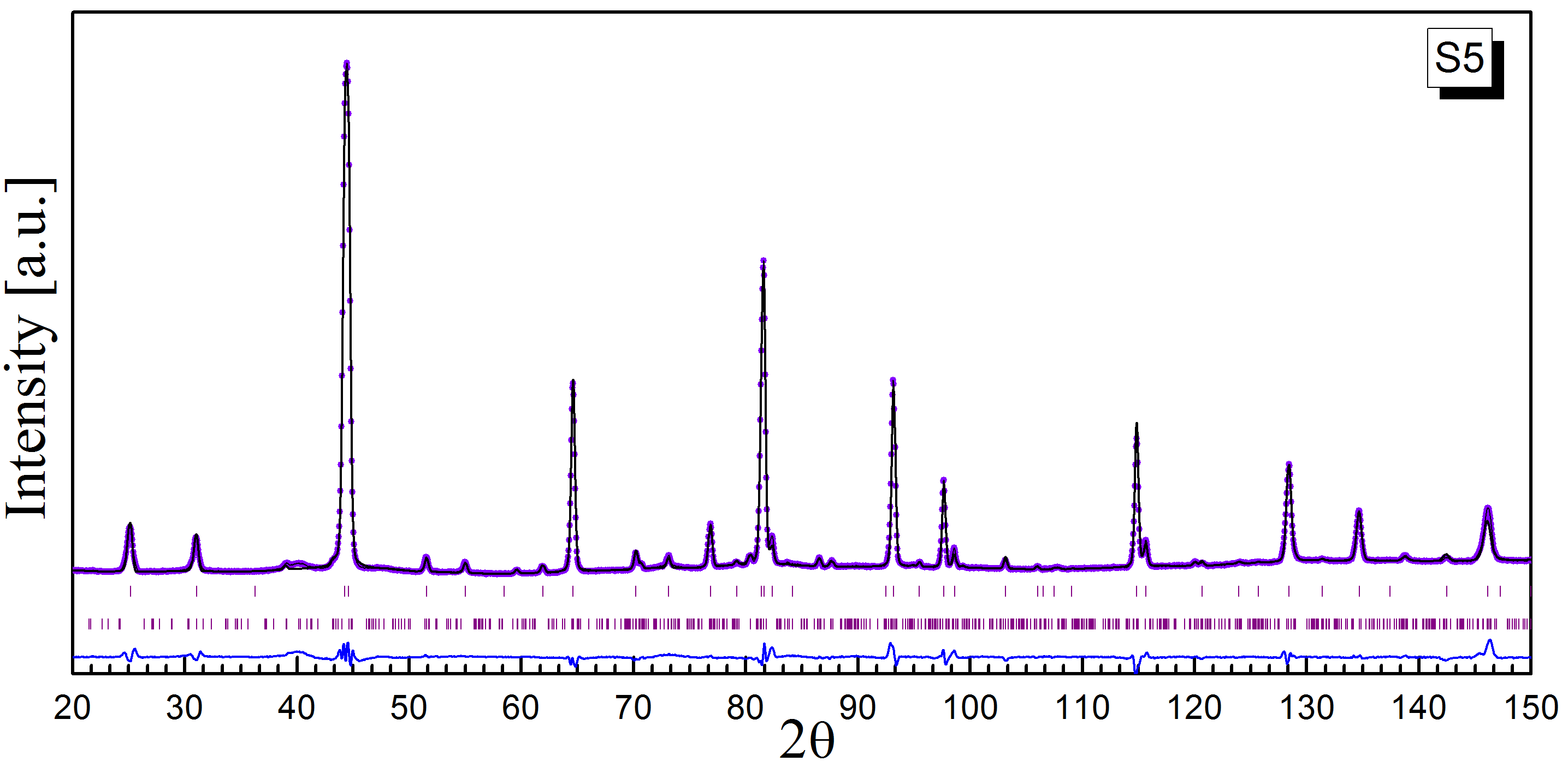}
\includegraphics[width=0.75\linewidth]{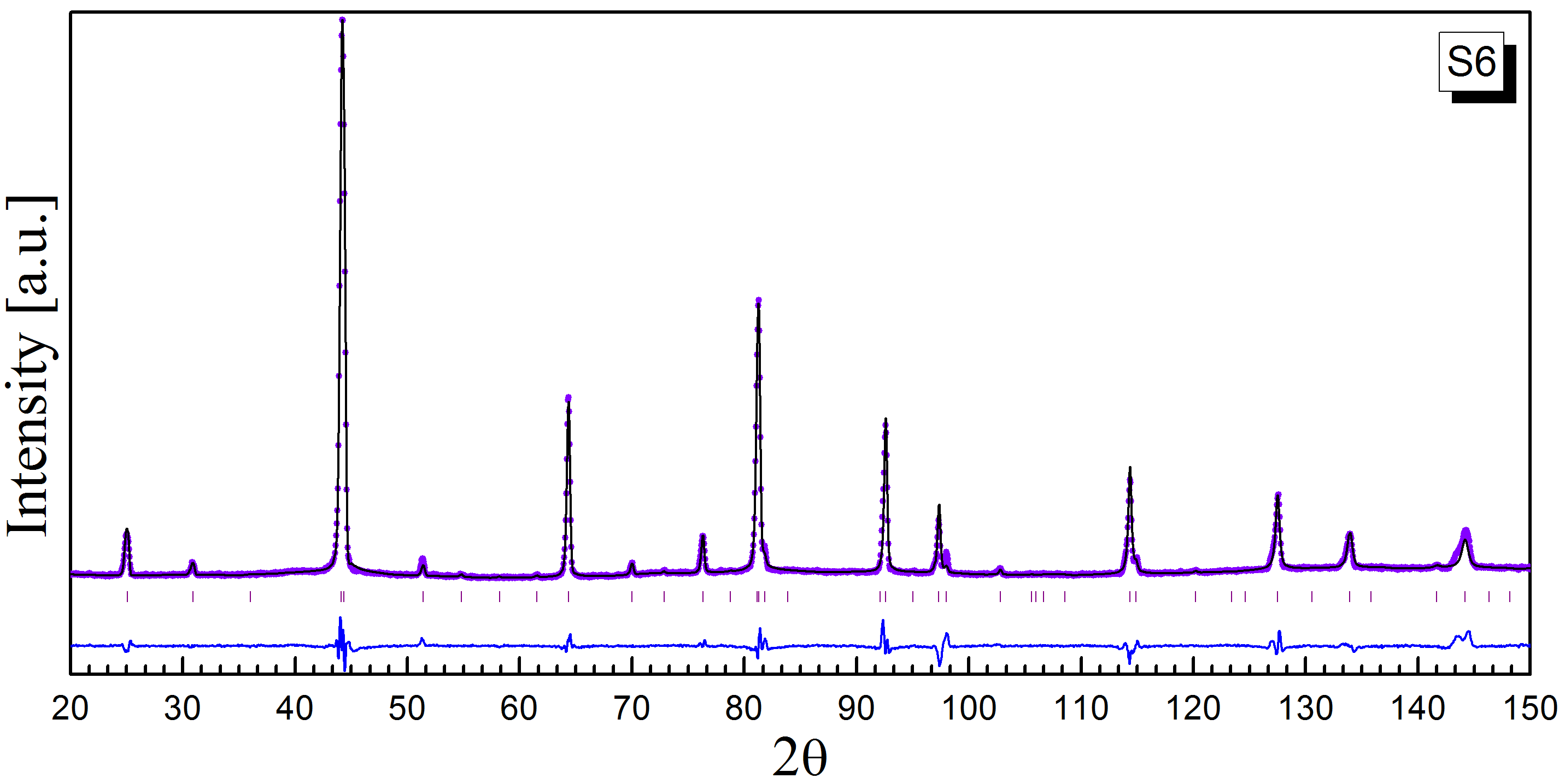}
\includegraphics[width=0.75\linewidth]{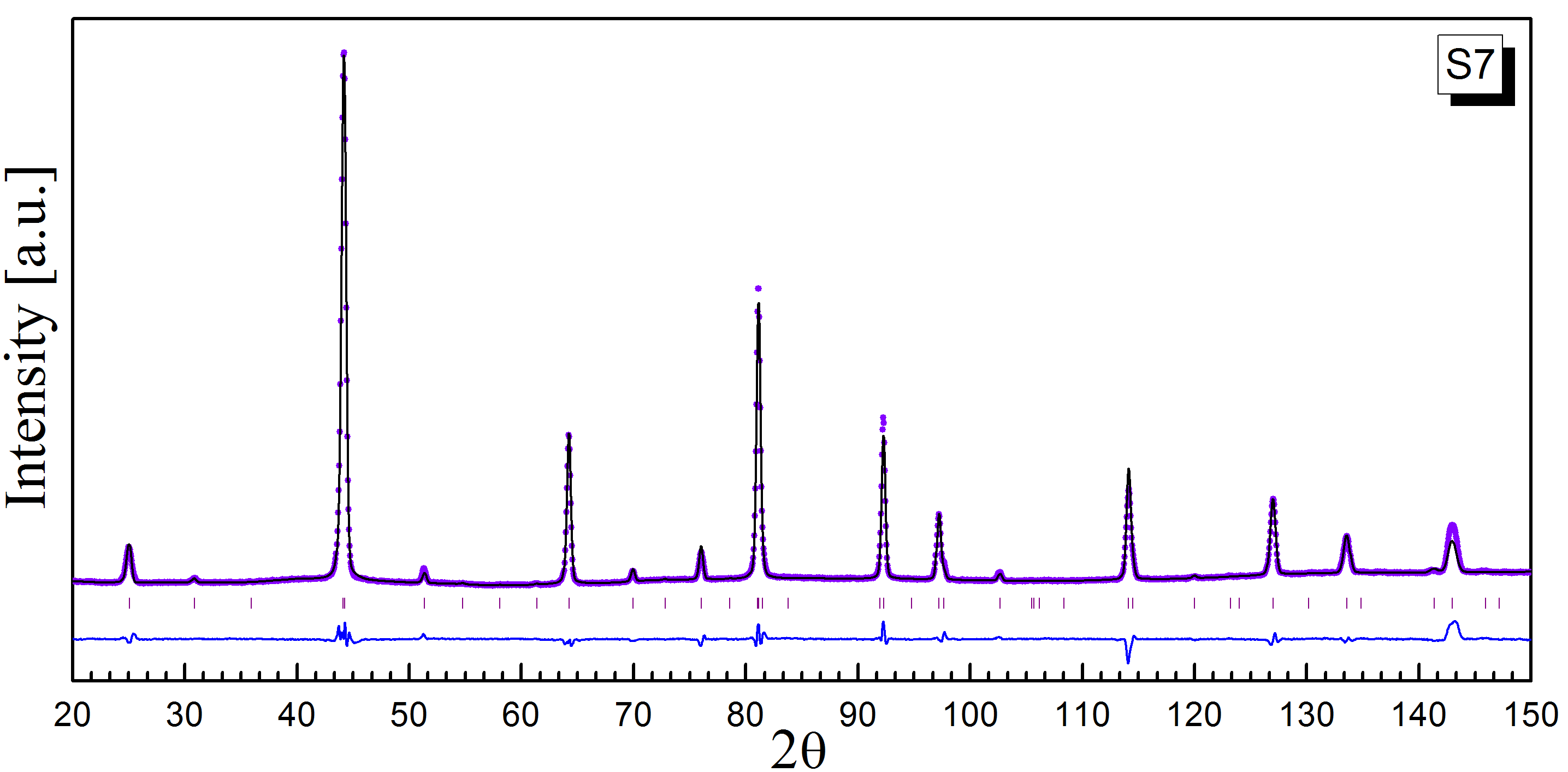}
\caption{Rietveld refinements for samples S5, S6 and S7 with nominal compositions: 58 at.\% Cu- 30 at.\% Sn- 12 at\% In, 60 at.\% Cu- 20 at.\% Sn- 20 at\% In and 60 at.\% Cu- 16 at.\% Sn- 24 at\% In, queched from $300{^{\circ}}$C, from data collected at room temperature in D2B. Vertical bars at
the bottom indicate Bragg reflections from the phases included in the
refinement: the HT-$\eta$-phase and the orthorhombic $\varepsilon$-phase (just in S5).The line at the bottom indicates the difference between the experimental and calculated patterns.} \label{f:refS5-S7}
\end{figure}

Figure\s\ref{f:refS5-S7} shows the ND data for samples S5, S6 and S7. When In atoms are added to the alloys, we can see from the diffraction data that no further reflections appear, indicating that In is indeed being incorporated in the crystal structure of Cu$_6$Sn$_5$. However, a further set of samples with increasing In content along the proposed $\eta$-phase field, which will be reported in a separate paper, already show the presence of additional reflections for 27 at.\% In\s\cite{tobe}. In the remaining we shall focus on samples with less than 25 at.\% In.

The question now arises about the location of the In atoms in the $\eta$-phase. In the Rietveld refinement, it is not possible to set all the occuppancies as free parameters given that there are three atomic species to allocate in three possible crystallographic sites, resulting in an undeterminated system of equations. A second set of independent experimental data would be needed to solve the problem, such as XRD. However, our attempts to collect XRD data not influenced by surface effects, lack of pollicrystallinity and grain size are still not reliable enough regarding the relative intensities of diffraction peaks, so they were not used to perform a multi-pattern refinement.
Instead, we selected among the various possibilities, two models which are consistent with the atomic distibution in both end-binaries. 
In model A, copper atoms completely occupy site \textit{2a}, as was unamibiguously shown by the refinement of sample S1. Tin and indium atoms completely occupy site \textit{2c} in proportions according to the stoichiometry of the alloy, and excess copper occupies site \textit{2d} which may remain partially unoccupied. In model B, site \textit{2c} is completely occupied by tin, whereas site \textit{2d} is shared by copper and indium atoms. Based on the above models and the nominal composition of each alloy, the occupancy at each site should  be ruled by the following relations
\begin{eqnarray*}
 \frac{1+Cu(2d)} {2+Cu(2d)+In(2d)}& = &  \mathrm{at.\%Cu} \\
 \frac{In(2c)+In(2d)} {2+Cu(2d)+In(2d)}  & =&   \mathrm{at.\%In} \\
 \frac{1-In(2c)} {2+Cu(2d)+In(2d)} & = &  \mathrm{at.\%Sn}, 
\end{eqnarray*}
where $Cu(2d)$ stands for the occupancy of Cu at the \textit{2d} site.
For model A, we assume $In(2d)=0$ and calculate the remaining nominal occupancies as
\begin{eqnarray*}
 Cu(2d) &=& \frac{2 \times \mathrm{at.\%Cu} - 1} {1-\mathrm{at.\% Cu}} \\
 In(2c) &=& \frac{\textrm{at.\% In}}{1-\mathrm{at.\%Cu}. } \\
 Sn(2c) &=& 1- In(2c)
\end{eqnarray*}
      
For model B, we set $In(2c)=0$ and calculate the nominal occupancies as
\begin{eqnarray*}
 Cu(2d) &=& \frac{1 - 2 \times \mathrm{at.\%Cu} - \mathrm{at.\%In} } {\mathrm{at.\% Sn}} \\
 In(2d) &=& \frac{\textrm{at.\% In}}{\mathrm{at.\%Sn} } \\
 Sn(2c) &=& 1.
\end{eqnarray*}

Refinements were first conducted using the above nominal occupancies as fixed parameters. The comparison was subtle for S2, but as from S3 it was clear that model B could not account for the data. In a further step, some occupancies were set as free parameters starting from model A, which resulted in the fits shown in Fig.\s\ref{f:refS5-S7}. The same procedure was applied using model B, but the refined occupancies moved to values with no physical consistency, whereas the use of model A converged to very reasonable values. In Table\s\ref{t:2} we list the refined occupancies. Next to the refined values, the nominal occupancies based on model A are quoted in brackets. Occupancies in itallics indicate fixed parameters.

There are certain issues to remark. As stated above, only an independent set of experimental data would allow a simultaneous refinement of all the occupancies. The present model leads to a good agreement with the data, but there may be other combinations that are also possible. For this reason, further experiments on the same samples are in progress to tackle this question using a selective technique such as synchrotron X-ray fine structure spectroscopy. Secondly, let us recall that a further hint about the probable site for In could be at the Cu-In side of the phase diagram. The literature on the binary Cu$_2$In $\eta$-phase, however, presents some discrepancies. Focussing just on the hexagonal HT form of this phase, sharing the same space group as Cu$_6$Sn$_5$, the In atoms are reported either in the \textit{2c} site (ICSD 102982 after ref.\s\cite{42Lav}, ICSD 627998 after ref.\s\cite{64Kal}) or in the \textit{2d} site (ICSD 657611 after ref.\s\cite{92Che}). The discussion above would give support to the former.

In Table\s\ref{t:2} we summarize the most relevant crystallographic parameters obtained from the refinement of samples S1 to S7. Although the site occupancies are constrained to the above picture, the structural parameters should not be severely affected. In the following section we discuss their composition dependence.

\subsection{Structural parameters as a function of composition}

\begin{figure}[tb]
\centering \vspace{3mm}
\includegraphics[width=\linewidth]{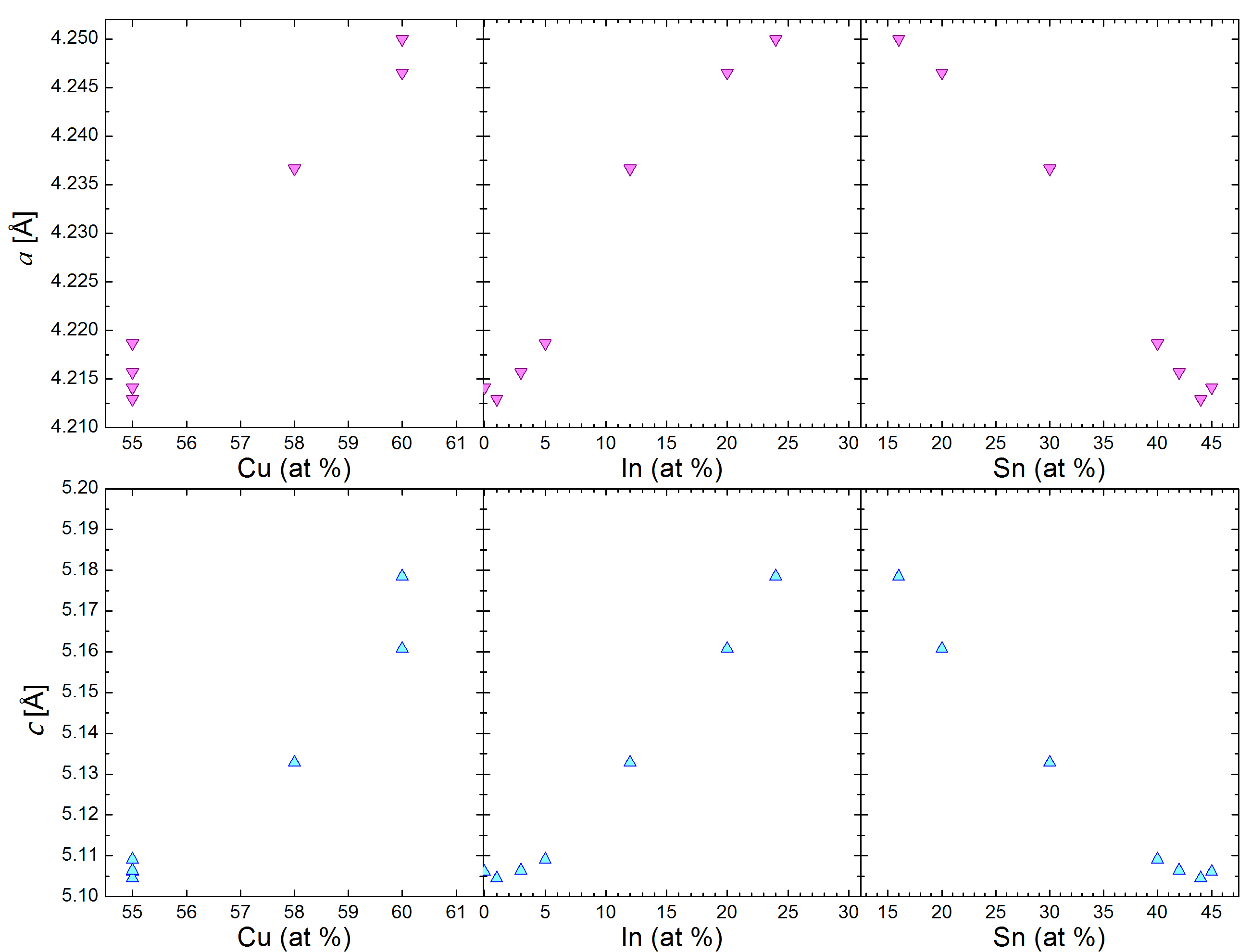}
\caption{Lattice parameters of the hexagonal $\eta$-phase as a function of each alloying element.} \label{f:LPsvsX}
\end{figure}

In Fig.\s\ref{f:LPsvsX} we present the lattice parameters $a$ (upper panels) and $c$ (lower panels) as a function of each component concentration. We observe that the $c$ parameter is controlled by the Cu concentration in the In-poor region, whereas for more In-concentrated alloys there is a linearity in the lattice parameters with concentration. 

\begin{figure}[tb]
\centering \vspace{3mm}
\includegraphics[width=0.75\linewidth]{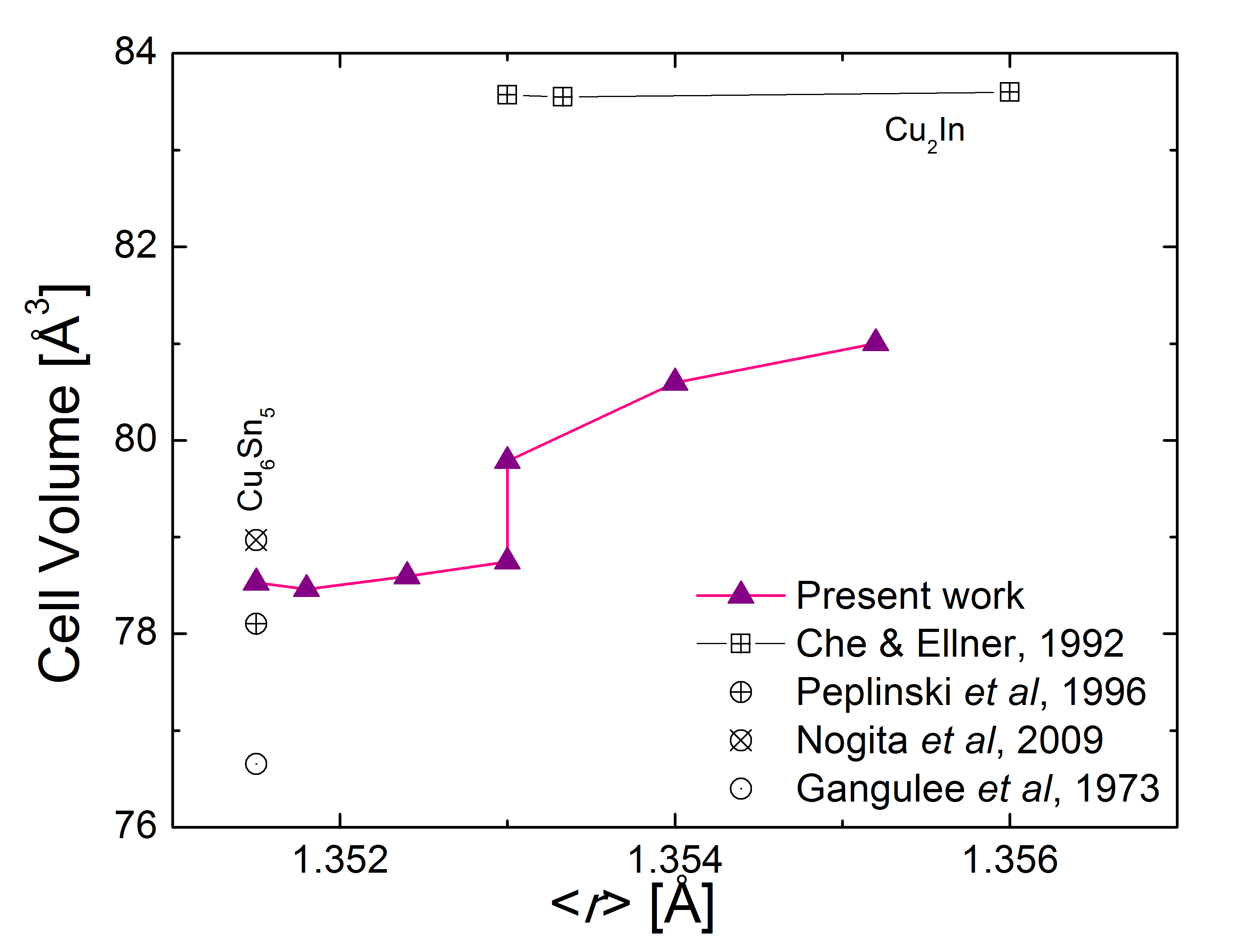}
\caption{Unit cell volume of the $\eta$-phase as a function of the average atomic (covalent) radius. Data for the binaries Cu$_6$Sn$_5$ and Cu$_2$In by \cite{96Pep,73Gan,09Nog,92Che} are also presented.} \label{f:Vegard}
\end{figure}

\begin{figure}[tb]
\centering \vspace{3mm}
\includegraphics[width=\linewidth]{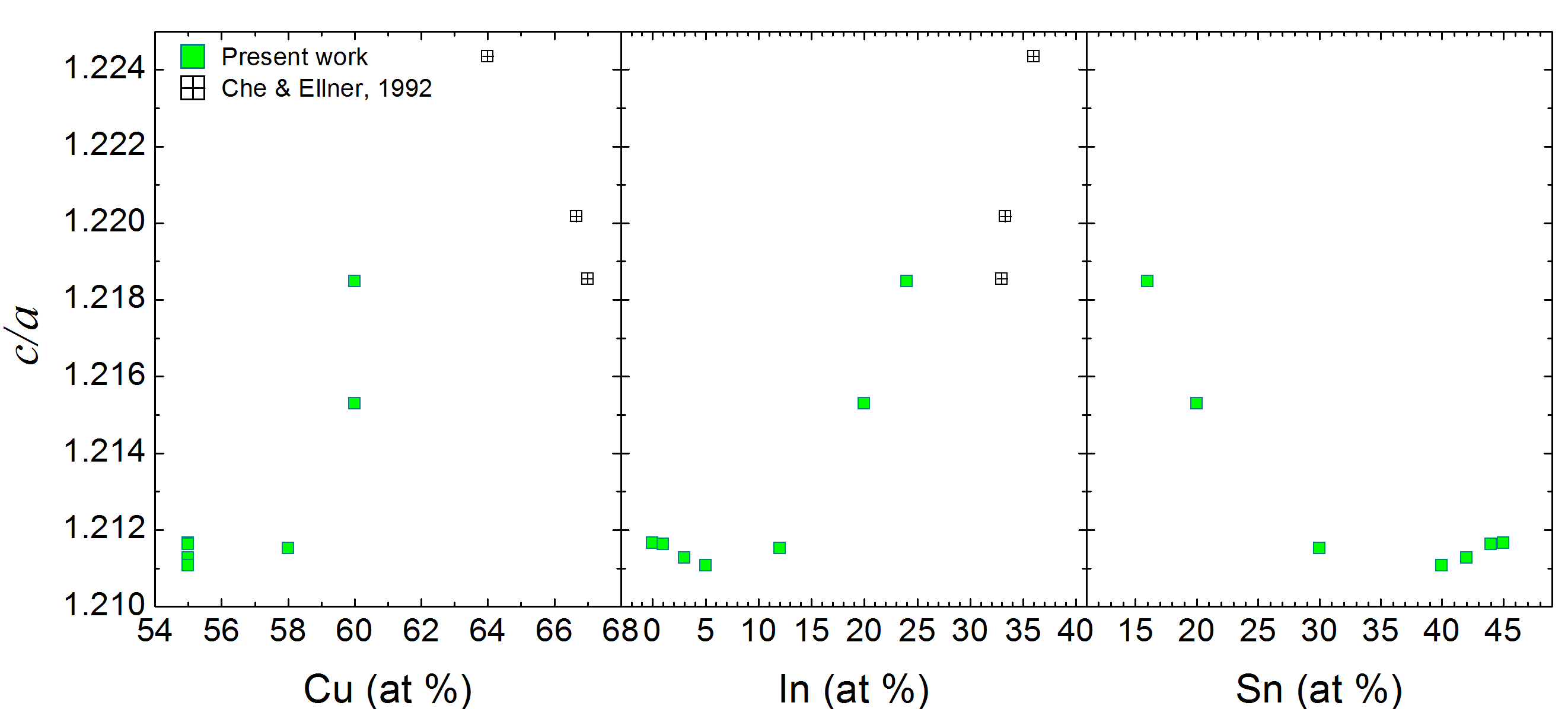}
\caption{Axial ratio $c/a$ for the ternary $\eta$-phase as a function of concentration of each alloying element.} \label{f:caratio}
\end{figure}

In order to rationalice the composition dependence of the $\eta-$phase volume, we present in Fig.\s\ref{f:Vegard} a plot of the unit cell volume as a function of the average atomic (covalent) radius, calculated as $<r>[\mathrm{\AA}]=\mathrm{at.\% Cu} \times {1.32 \mathrm{ \AA}} + \mathrm{at.\% Sn} \times {1.39 \mathrm{ \AA}}+\mathrm{at.\% In} \times {1.42 \mathrm{ \AA}}$ where the values for the individual atomic species were taken from Ref.\s\cite{08Cor}. On the same graph we have represented the calculated volume for data from the binaries Cu$_6$Sn$_5$ and Cu$_2$In. It is interesting to note that for the first four alloys, sharing the same Cu content, the volume of the cell remains almost unchanged, and the volume is governed by the Cu fraction. Between samples S4 and S5, although they have the same average radius $<r>$, there is a jump in volume due most probably to the doubling of the In content together with an increase in Cu from 55 at.\% to 58 at.\%. The alloys reported by Che and Ellner\s\cite{92Che} on the Cu$_2$In phase, with a Cu concentration around 66 at.\%, illustrate clearly the strong dependece of volume with Cu concentration. Finally, we show in Fig.\s\ref{f:caratio} the $c/a$ ratio for the present alloys, together with that reported by Che and Ellner\s\cite{92Che} for Cu$_2$In.

\section{Final remarks}

Although much work has been devoted to the study of intermetallic phases in Cu-Sn and Cu-In alloys, a systematic diffraction study of ternary Cu-In-Sn alloys was still lacking. The present work has been devoted to study the $\eta$-phase field region in the ternary phase diagram. We present here neutron diffraction data, a technique not used before --to the best of our knowledge-- to study these systems. We found this technique to be extremely useful to overcome common experimental difficulties related to XRD measurements, which may be at the origin of the confusing and scattered information available in the databases.   
Our results confirm that in the range 0-25 at.\% In addition to the Cu$_6$Sn$_5$ base alloy, the $\eta$-phase can stabilize at 300$^{\circ}$C without forming further superstructures, as proposed in the available phase diagrams\s\cite{72Koe,09Lin}. However, further In addition results in the appearence of a number of extra, weak Bragg reflections indicating either a phase coexistence or a superstructure formation. Those alloys are not discussed in the present paper, and will be reported separately. A discussion has also been presented about the possible site occupancies for each element in the ternary phase. 
To sum up, structural data for the ternary $\eta$- phase are reported, which can be of value in the study and modelisation of diffusion processes for new bonding technologies, as well as for calculations of stable and metastable phases in the ternary phase diagram.


\begin{acknowledgements}
This work is part of a research project supported by Agencia Nacional de Promoci\'on Cient\'ifica y Tecnol\'ogica, under grant PICT2006-1947. We particularly acknowledge ILL and its staff for
the beamtime allocation and technical assistance. 
\end{acknowledgements}

\bibliographystyle{spphys}       
\bibliography{/home/gaby/Documents/Cu-Sn-In/biblio/Cu-Sn-In}

\begin{thebibliography}{10}
\providecommand{\url}[1]{{#1}}
\providecommand{\urlprefix}{URL }
\expandafter\ifx\csname urlstyle\endcsname\relax
  \providecommand{\doi}[1]{DOI \discretionary{}{}{}#1}\else
  \providecommand{\doi}{DOI \discretionary{}{}{}\begingroup
  \urlstyle{rm}\Url}\fi

\bibitem{08Din}
A.T. Dinsdale, A.~Kroupa, J.~V{\'i}zdal, J.~Vrestal, A.~Watson, A.~Zemanova,
  COST Action 531-Atlas of Phase Diagrams for Lead-free Solders \textbf{1}, 182
  (2008)

\bibitem{04Gal}
W.~Gale, D.~Butts, Science and Technology of Welding \& Joining \textbf{9}, 283
  (2004).
\newblock \doi{doi:10.1179/136217104225021724}

\bibitem{07Vel}
T.~Velikanova, M.~Turchanin, O.~Fabrichnaya, \emph{Cu-In-Sn
  (Copper-Indium-Tin)} (Materials Science International Team MSIT®, 70507
  Stuttgart, Germany, 2007), p. 249.
\newblock Non-Ferrous Metal Ternary Systems. Selected Soldering and Brazing
  Systems: Phase Diagrams, Crystallographic and Thermodynamic Data - New Series
  IV/11C3

\bibitem{72Koe}
W.~Koester, T.~Goedecke, D.~Heine, Zeitschrift fuer Metallkunde
  \textbf{63}(12), 802 (1972)

\bibitem{08Lin}
S.K. Lin, C.F. Yang, S.H. Wu, S.W. Chen, Journal of Electronic Materials
  \textbf{37}, 498 (2008).
\newblock 10.1007/s11664-008-0380-0

\bibitem{09Lin}
S.K. Lin, T.Y. Chung, S.W. Chen, C.~horng Chang, Journal of Materials Research
  \textbf{24}(08), 2628 (2009).
\newblock \doi{10.1557/jmr.2009.0317}

\bibitem{01Liu}
X.J. Liu, H.S. Liu, I.~Ohnuma, R.~Kainuma, K.~Ishida, S.~Itabashi, K.~Kameda,
  K.~Yamaguchi, Journal of Electronic Materials \textbf{30}(9), 1093 (2001)

\bibitem{97Eld}
M.~Elding-Pont{\'e}n, L.~Stenberg, S.~Lidin, Journal of Alloys and Compounds
  \textbf{261}(1-2), 162 (1997)

\bibitem{03Bah}
Z.~Bahari, E.~Dichi, B.~Legendre, J.~Dugu{\'e}, Thermochimica Acta
  \textbf{401}(2), 131 (2003)

\bibitem{72Jai}
K.C. Jain, M.~Ellner, K.~Schubert, Zeitschrift fuer Metallkunde \textbf{63},
  258 (1972)

\bibitem{89Sub}
P.R. Subramanian, D.E. Laughlin, Bulletin of Alloy Phase Diagrams
  \textbf{10}(5), 554 (1989)

\bibitem{10Nog}
K.~Nogita, Intermetallics \textbf{18}(1), 145 (2010).
\newblock \doi{DOI: 10.1016/j.intermet.2009.07.005}

\bibitem{10Sch}
U.~Schwingenschl{\"o}gl, C.D. Paola, K.~Nogita, C.M. Gourlay, Applied Physics
  Letters \textbf{96}(6), 061908 (2010).
\newblock \doi{10.1063/1.3310019}

\bibitem{96Pep}
B.~Peplinski, G.~Schulz, D.~Schultze, E.~Schierhornand, Materials Science Forum
  \textbf{228-231}, 577 (1996).
\newblock \doi{10.4028/www.scientific.net/MSF.228-231.577}

\bibitem{90Mas}
T.B. Massalski, Binary Alloy Phase Diagrams 2nd. Edition  (1990)

\bibitem{fullprofB}
J.~Rodr{\'i}guez-Carvajal, in \emph{Abstracts of the Satellite Meeting on
  Powder Diffraction of the XV Congress of the IUCr, Toulouse, France} (1990),
  p. 127

\bibitem{73Gan}
A.~Gangulee, G.C. Das, M.B. Bever, Metallurgical and Materials Transactions B
  \textbf{4}, 2063 (1973).
\newblock 10.1007/BF02643268

\bibitem{42Lav}
F.~Laves, H.J. Wallbaum, Zeitschrift f{\"u}r Anorganische und Allgemeine Chemie
  \textbf{250}(1), 110 (1942).
\newblock \doi{10.1002/zaac.19422500110}

\bibitem{64Kal}
R.S. Kalyana-Raman, R.K. Gupta, M.N. Sujir, S.~Bhan, Journal of Scientific
  Research of the Banaras Hindu University \textbf{14}, 95 (1964)

\bibitem{09Nog}
K.~Nogita, C.~Gourlay, T.~Nishimura, JOM Journal of the Minerals, Metals and
  Materials Society \textbf{61}, 45 (2009).
\newblock 10.1007/s11837-009-0087-6

\bibitem{83Wat}
Y.~Watanabe, Y.~Fujinaga, H.~Iwasaki, Acta Crystallographica Section B
  \textbf{39}(3), 306 (1983).
\newblock \doi{10.1107/S0108768183002451}

\bibitem{70Bro}
P.~Brooks, E.~Gillam, Acta Metallurgica \textbf{18}(11), 1181 (1970).
\newblock \doi{DOI: 10.1016/0001-6160(70)90108-2}

\bibitem{tobe}
G.~Aurelio, S.A. Sommadossi, G.J. Cuello, {Neutron diffraction study of the
  stability and phase transitions in Cu-Sn-In alloys as alternative Pb-free
  solders} (2012).
\newblock To be published

\bibitem{VESTA}
K.~Momma, F.~Izumi, Journal of Applied Crystallography \textbf{41}(3), 653
  (2008).
\newblock \doi{10.1107/S0021889808012016}

\bibitem{92Che}
G.C. Che, M.~Ellner, Powder Diffraction \textbf{7}(2), 107 (1992)

\bibitem{08Cor}
B.~Cordero, V.~G{\'o}mez, A.E. Platero-Prats, M.~Reves, J.~Echeverr{\'i}a,
  E.~Cremades, F.~Barragan, S.~{\'A}lvarez, Dalton Trans. pp. 2832--2838
  (2008).
\newblock \doi{10.1039/B801115J}

\end{thebibliography}

%
%

\end{document}